# Application of a design space exploration tool to enhance interleaver generation

CHAVET Cyrille[1], COUSSY Philippe[2], URARD Pascal[1], MARTIN Eric[2]
[1]STMicroelectronics, Crolles, FRANCE. {firstname.lastname@st.com}
[2]LESTER Lab, UBS University, CNRS FRE 2734. {firstname.lastname@univ-ubs.fr}

**ABSTRACT**

*This paper presents a methodology to efficiently explore the design space of communication adapters. In most digital signal processing (DSP) applications, the overall performance of the system is significantly affected by communication architectures, as a consequence the designers need specifically optimized adapters. By explicitly modeling these communications within an effective graph-theoretic model and analysis framework, we automatically generate an optimized architecture, named Space-Time AdapteR (STAR). Our design flow inputs a C description of Input/Output data scheduling, and user requirements (throughput, latency, parallelism…), and formalizes communication constraints through a Resource Constraints Graph (RCG). Design space exploration is then performed through associated tools, to synthesize a STAR component under time-to-market constraints. The proposed approach has been tested to design an industrial data mixing block example: an Ultra-Wideband interleaver.*

## 1. INTRODUCTION

In the multimedia and telecommunications domain, continuously emerging customer services require severe performance (computing power, timing performances and memory bandwidth/capacity) to implement the new communication standards. Indeed, communication system applications require high throughput -on the order of several hundred Mb/s- accompanied by both low latency and severe bit error rate BER constraints (e.g. wireless, fiber-optic communication…). Owing to their impressive near-Shannon-limit error correcting performance, turbo-like codes in their parallel or serially concatenated versions [2], originally dedicated to channel coding, are being currently reused in a large set of the whole digital communication systems (e.g. equalization, demodulation, synchronization, MIMO).

These codes are formed by two or more processing elements PE (encoders/decoders) and one communication network that interleaves the data blocks exchanged by the PEs. The turbo decoding principle is based on an iterative algorithm using decoders exchanging information in order to improve the error correction performance through the iterations. The iterative nature of these algorithms is a severe constraint to satisfy the aforementioned requirements with an affordable implementation complexity. A widespread solution is to realize the turbo decoder in a parallel fashion. One the one hand, this solution increases the throughput since the latency of the system becomes the latency of constituent sub-blocks [6]. On the other hand, the complexity and the cost of the system are increased due to parallel nature of the architecture. Moreover, for the sub-blocks to be able to work in parallel, it is necessary that each one exchanges data with a Random Access Memory block (RAM).

By the way, depending on the specific permutation law, different modules may try to simultaneously access the same RAM. As a consequence, none of them is able to retrieve data. This problem is known as the "collision" problem [9]. In this case, the access to the memory has to be postponed and carefully arbitrated, which slows down the decoding process. The solution consists in designing an adapted interleaver and/or modifying the decoder architecture. In this paper, we propose to use the formal approach presented in [3] to tackle the interleaver design problem.

The paper is organized as follows: the second section presents a state-of-the-art for interleaving architectures. The third section is dedicated to the problem formulation of the interleaver design. In the fourth section we briefly present our design flow. Finally, the last section presents experimental results and the design exploration offered by our design flow on an industrial example.

## 2. INTERLEAVING ARCHITECTURE

Interleaving is a permutation rule that scrambles data to break up neighbourhood-relations. It is a key factor for turbo-codes performances which varies from one standard to another. Moreover within a given standard, different interleaving rules can be used for different modes through varying frame lengths and/or data rates. In this context, taking into account the aforementioned constraints and the collision problems, hardware implementations of parallel turbo decoders require the integration of complex topology supporting the intensive interleaved memory accesses. Indeed, in state-of-the-art parallel turbo-decoding, interleaving is considered as a limiting factor concerning the overall system performance and the architectural cost.

To successfully tackle these problems, different solutions have been recently proposed. First, possible solutions to get rid of collisions with nonprunable interleavers, consist in designing a specific interleaver rule. In [9], the authors propose a deterministic methodology to design collision-free interleavers. In [10] and [8] the authors define collision-free permutations thanks to a combination of a spatial and a temporal permutation. The authors of [12] simply integrate the collision-free constraint in the design of their interleaver. However, the multi-modes architectures (depending on frame length, data-rate…) can not be handled by such approaches. Another solution consists in defining a collision-free interleaver that preserves this property even when pruned. In [7], the authors describe a design rule to obtain such interleavers, with an incremental algorithm that generates collision-free interleavers by adding new elements in successive steps to a small permutation. Of course, all these solutions are viable only if the designer is free to choose the permutation law to be used in the system. As a consequence, the resulting architecture may not be standard compliant.

In [16] the authors propose, in case of a collision, to store the conflicting information in the communication network until the targeted sub-block can process it. Of course, additional network buffering resources, and consequently time needed to interleave information, increase with the number of parallel processors. This is a suboptimal strategy, in terms of latency and thus throughput, which avoids collisions at the expense of area and memory. Moreover, the communication is based on a Benes network [5], which might be suboptimal compared to a dedicated and optimized architecture.

Unlike these implementations, in [13] the authors propose a solution based on software and/or reconfigurable parts to achieve the required flexibility, but achieving lower throughput. In [14], an advanced heterogeneous communication network implementation was proposed. Two multistage interconnection network architectures are presented in order to handle on-chip communications in multiprocessor parallel turbo decoders. They are based on a dedicated network and associated routers. The main feature of these network architectures (Butterfly and Benes based topologies) is their supposed scalability enabling seamless trade-off between hardware complexity and available bandwidth for turbo decoding. The Butterfly network, which lacks of diversity, is a multistage interconnection network with 2-input 2-output routers. There is a unique path between each source and destination. As a consequence, the risk of conflict is increased and the authors have to add queues to store conflicting information. The second network architecture proposed is based on a Benes network. In this case, the latency is constant for all the couples (source, destination), but this network avoids the conflicts if and only if all the paths have a different destination. Unfortunately, we saw that it was not true for turbo-decoding applications because interleaving (respectively de-interleaving) ends in potential conflicts. Moreover, as already mentioned the Benes networks are costly and under-optimized solutions.

Finally, the authors of [15] describe a system that avoids collisions for every interleaver and any degree of parallelism. This solution consists in automatically finding a collision-free data memory mapping respecting the interleaving rule, thanks to a simulated-annealing algorithm. As a consequence, the user cannot predict when the algorithm will end. Moreover, the proposed approach does not target the optimization of the storage elements.

In this paper, we propose to use the formal approach presented in [3] to tackle the interleaver design problem. This approach, which originally target interface synthesis, is shown to be also suited to the interleaver design space exploration. Our design flow can take as input timing diagrams (constraints file) or C descriptions of I/O data scheduling (e.g. an interleaving formula), with user requirements (throughput, latency…). We formalize communication constraints through a formal Resource Constraints Graph (RCG) which properties enable an efficient architecture exploration. By using our design flow, any user can generate an optimized architecture in term of latency, network architecture and memory, from any interleaving standard.

## 3. PROBLEM FORMULATION

First, from throughput and parallelism constraints, and an interleaver permutation pattern, we can formalize data communication as timing diagrams.

Let us consider a simple architecture example composed of two components exchanging a set of data $S = \{a, b, c, d, e, f\}$. S is produced by a block #1 and is consumed by a block #2 through a single point-to-point link. The write access sequence into the communication link is $S_w = (a,c,b,e,f,d)$ i.e. $t^w_a < t^w_c < t^w_b < t^w_e < t^w_f < t^w_d$, while the read access sequence from the link is different $S_r = (c,a,e,b,d,f)$ i.e. $t^r_c < t^r_a < t^r_e < t^r_b < t^r_d < t^r_f$ (see Figure 1).

This difference between the two I/O sequences can either come from the integration of two IP cores that were not specifically designed to work together, either can be explicitly described (e.g. in interleavers [11][7]). Those blocks may not produce and consume data in the same order nor with the same throughput (nor sometime the same parallelism), so they can not be directly plugged together. The designer needs to introduce a space-time adapter between them to ensure correct functional results. A classical solution consists in using a memory to buffer all concerned data: this is what we call coarse grain approach. But in fact, this over sized buffer may be reduced thanks to a finer grain communication constraints analysis [1]. The proposed adapter can be designed either by using a set of registers or specific memory elements, such as FIFO (queue) or LIFO (stack). The problem the designer faces consists in finding the best architecture for this adapter: he has to find the best storage element binding in order to integrate data reordering and to minimize total amount of memory.

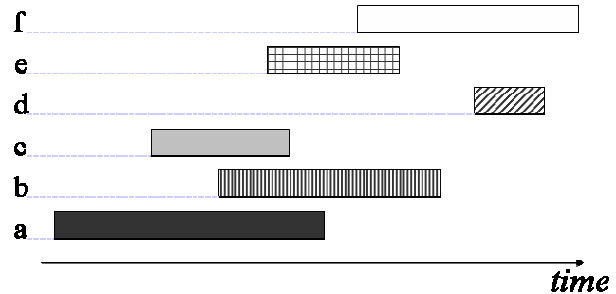

Figure 1: Data timing diagram.

For example, the lifetimes of data $a$ and $b$ respect a First-In First-Out semantic, so they can be assigned to the same hardware FIFO. This timing relation is also true for the data $c$ and $b$. However, data $a$ and $c$ respect a Last-In First-Out semantic, so a single hardware FIFO cannot be used to store the data $a$, $b$ and $c$ The question for the designer is: how can we bind data $a$, $b$ and $c$ to different storage elements, in order to generate the best final architecture? This highlights the fact that the local problem of $a$, $b$ and $c$ binding will influence the resulting global architecture. A methodology is thus needed to bind data $a$, $b$ and $c$ to different storage elements, in order to generate an optimized architecture.

In a nutshell, a designer needs (1) a tool to generate timing diagram from interleaving permutation scheme and architectural requirements (e.g. parallelism), (2) a tool to generate the corresponding architecture and (3) a tool to validate this architecture.

## 4. STAR DESIGN FLOW

The architecture of a STAR component is composed of a datapath and the associated control state machine FSM (see Figure 2). The data path can be composed of FIFO, LIFO or register. Spatial adaptation (a data read on one input port can be send to any/several output ports) is performed by an interconnection logic dealing with data dispatching from input port to storage elements, and from storage elements to output ports. We can see on Figure 2 that there is one STAR architecture for each input port.

The timing adaptation (data-rates, different input/output data scheduling) is realized by the storage elements. STAR can have a GALS (Globally Asynchronous Locally Synchronous) / LIS (Latency Insensitive System) using the mechanisms described in [1].

The design flow is presented in Figure 4 and is based on three tools: *StarTor* for the STAR design constraint specification, *StarGene* for the STAR component synthesis and *StarBench* for the STAR functional validation. The methodology generates a register transfer level (RTL) VHDL architecture, and associated DC Synopsys scripts, starting from a functional model and a set of user requirements (timing and communication-architecture constraints).

The architecture synthesis is performed by using a library of pre-designed and characterized storage elements (FIFO, LIFO and Registers).

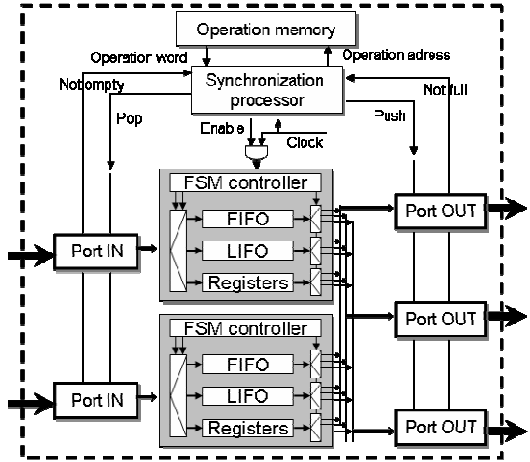

Figure 2: Typical STAR architecture.

*StarTor* inputs a C level algorithmic description which specifies the interleaving scheme, and a file containing user requirements (latency, throughput, communication interface, I/O parallelism...). StarTor first extracts I/O data communication order by generating a trace from the execution of the C functional description. Next, based on designer's requirements, it generates a constraints file. This file contains the number and type of ports, type and amount of data, relationships between data and ports (i.e. mapping) and finally read and write access dates for all data. The design can generate a set of architectural constraints to compare one to each other.

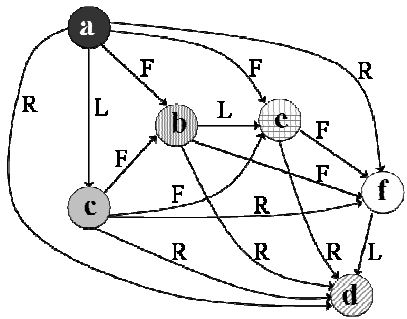

Figure 3: Graph example (from Figure 1 constraints).

Then, in order to generate a STAR component, our design tool *STARGene* is based on a four-step flow: (1) Resource Compatibility Graph construction, (2) Storage resource binding, (3) Architecture optimization and (4) VHDL RTL generation. During the first step of the STAR component Generation, a Resource Constraints Graph RCG (see Figure 3) is generated from the communication constraints. The analysis of this formal model allows both data binding to storage elements (queue, stack or register), and the sizing of each storage element. This first architecture is next optimized by merging storage elements that have non-overlapping usage timing frames. Finally, an RTL level design is generated.

The last tool, *StarBench*, generates a test bench based on constraints in order to validate the design by comparing simulation results.

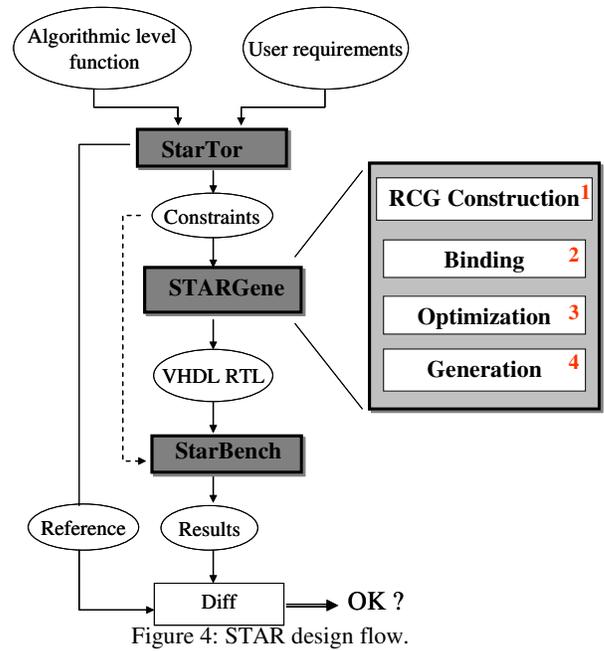

Figure 4: STAR design flow.

The design space exploration is driven by the designer which can plot a set of exploration parameters. The main ones are:

- *Enabling FIFO/LIFO/REG*: the user may switch off FIFO and/or LIFO and/or Register binding on the RCG.

- *FIFO/LIFO minimal/maximal size*: in order to avoid parasitic structures like too small or too big FIFO/LIFO structures (structure size). Using this parameter, the user is able to limit the size of the generated FIFO/LIFO structure to meet its own technological constraints (e.g. maximum size of 1024 elements).

- *FIFO/LIFO average use factor*: the user can define a minimal usage for the FIFO/LIFO structures to be binded. This metric aims to count the total amount of data travelling in each memory element during an iteration of the interleaving algorithm, compared to its size (e.g. 274 elements in a 64-places FIFO).

- *FIFO/LIFO filling factor*: the user can define a minimal number of data in a FIFO/LIFO. The generated FIFO/LIFO sizes are power of 2, so if the maximum number of data in a given FIFO at the same time is 52, we generate a 64-places FIFO. This factor aims to limit the gap between the depth of the generated FIFO and the maximum data that will be stored in it at a given time.

- *Multiplexer complexity factor*: a STAR may have to deal with high parallelism architectures. Thus, a given structure (FIFO, LIFO or register) may have to store data from multiple inputs or send data to multiple output ports. In order to avoid the generation of a complex communication network, the exploration algorithms is also driven by a dedicated metric which aims at reducing this complexity.

- *Weighting each parameter*: Each of the previous parameters can be balanced by the user by means of dedicated coefficients. During the RCG exploration, the binding algorithm explores the graph, and binds structures thanks to these metrics.

**Table 1:** Design space exploration by means of plotted metrics

| # Data | IN par. | OUT par. | Mux Factor | # min. | Usage factor | Structure binding | | # structures | | | | | # memory points | Throughput |
|---|---|---|---|---|---|---|---|---|---|---|---|---|---|---|
| | | | | | | FIFO | LIFO | *Mux* | *FIFO* | *LIFO* | *Register* | **Total** | | |
| 600 | 6 | 5 | Yes | 4 | 90% | Yes | Yes | 48 | 45 | 5 | 22 | 174 | 514 | 272,7 |
| 600 | 6 | 5 | Yes | 7 | 90% | Yes | Yes | 34 | 34 | 1 | 90 | 159 | 525 | 272,7 |
| 600 | 6 | 5 | Yes | 15 | 90% | Yes | Yes | 12 | 12 | 0 | 391 | 415 | 581 | 272,7 |
| 600 | 6 | 5 | Yes | 30 | 90% | Yes | Yes | 0 | 0 | 0 | 600 | 600 | 600 | 272,7 |
| 600 | 6 | 5 | Yes | 7 | 50% | Yes | Yes | 28 | 27 | 2 | 73 | 130 | 533 | 272,7 |
| 600 | 6 | 5 | Yes | 7 | 70% | Yes | Yes | 29 | 28 | 2 | 70 | 129 | 532 | 272,7 |
| 600 | 6 | 5 | Yes | 7 | 80% | Yes | Yes | 33 | 32 | 1 | 78 | 144 | 534 | 272,7 |
| 600 | 6 | 5 | Yes | 7 | 100% | Yes | Yes | 39 | 38 | 1 | 84 | 162 | 522 | 272,7 |

In field of turbo-like architecture, communication traffic profile, which depends on the interleaving rule, may have to support multi-modes and multi-standard features. In this context, hardware implementations of parallel turbo decoders require the integration of complex topology and routing resources supporting the intensive interleaved memory accesses. The STAR design flow integrates design exploration for multi-modes architecture, switching from one to another at run-time. By the way, in this paper we present a formal methodology to synthesize a STAR architecture for a given configuration. Due to space limitation, the generalization of the methodology generating multi-mode architecture (graph merging, multi data path synthesis, multi FSM generation…) will be fully presented in a future publication.

### 4. ULTRA-WIDE BAND INTERLEAVER FOR STMICROELECTRONICS

In this section, we present the exploration of a STAR architecture on an industrial example. The application, an Ultra Wide Band interleaver [11], was provided by STMicroelectronics. This interleaver has to be able to switch between different modes (300, 600 or 1200 data length), respecting latency constraints. By nature, interleavers may offer few storage elements to be saved. However, these data-mixing schemes are well-suited for our proposed design flow and we can explore how metrics (I/O parallelism, enable/disable FIFO/LIFO, average usage factor…) may influence the final architecture. All the areas information has been masked in order to protect internal company technologies.

First, we present the interleaving law. Then we highlight the parallelism effect on the STAR architecture and we compare it to classical solutions. Finally, we show how the memory-related and the network-related metrics impact the resulting architecture.

#### 4.1 Parallelism exploration

Thanks to the *StarTor* tool, we efficiently explore how the parallelism influences the resulting architecture. Figure 5 shows the results for an architectural exploration with different I/O parallelism. The two reference architectures are based on either RAM block or on a sea of registers. The X-axis represent the parallelism (e.g. 4 means 4 data input and 4 data output). The Y-axis represents the number of memory points needed for the RAM based architecture, the sea of register and the STAR based architecture.

The number of memory points for the RAM based architecture is defined by the number needed RAM blocks (i.e. same as the input parallelism) multiplied by the number of memory points in each one (i.e. for 1000 8-bit words, a 1024 byte RAM block is needed). The STAR architecture needs less memory points than the RAM based architecture thanks to it FIFO/LIFO/register elements and the results are close to a register-only architecture.

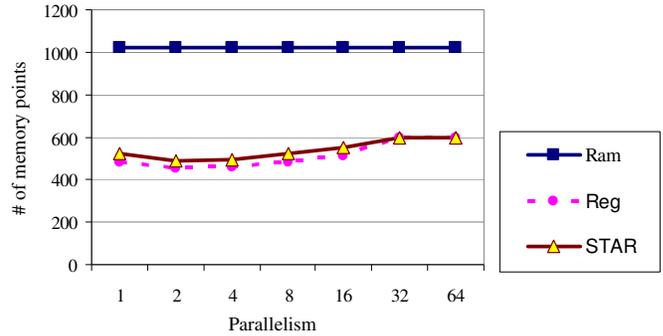

Figure 5: Exploration for different I/O parallelism

Figure 6 compare the overall complexity of the obtained architecture in term of control and communication network. For this purpose, we compared the number of structure to be controlled. This number is given by the number of memorizing elements plus the number of switches or multiplexer of the communication network.

In case of a RAM-based architecture, the communication network is based on a cross-bar in which, if there are N ports to be connected to N RAM, the network needs $N*N$ 2x2-switches to be piloted. Figure 6 shows that for low parallelism the RAM-based architecture can be better than the STAR.

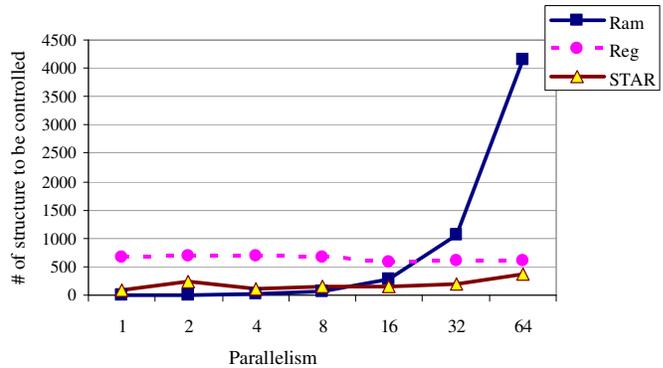

Figure 6: Architecture complexity in term of # of structures to control

However, the STAR architecture is far less complex for higher parallelisms which are widely used in current design.
Moreover, the Register-based architecture is always more complex than the STAR architecture in terms of structure to be controlled.

## 4.2 Design space exploration through plotted metrics

4.2.1 Memory related metrics

In table 1, we show the resulting architectures for a given I/O parallelism while exploring two metrics: *structure minimal size* and *structure usage factor*. In this table, the first column indicates the frame length (600 data in this example), while the columns "*IN* and *OUT par.*" indicate the used parallelism. The columns "*Mux factor*", "*# min*", "*Usage factor*" refer to the corresponding parameters (e.g. # min = 7, means that the minimal size for a FIFO or a LIFO structure to be bind is 7) and the column "*Structure binding* " indicates if the binding of FIFO or LIFO are allowed during the exploration. The other column indicate the results: the number of FIFO, LIFO, Register and multiplexer, the number memory points needed and the throughput of the resulting architecture (in Mb./s.).

This experiment shows that these parameters greatly impact the architecture: e.g. the total number of structures to be controlled varies from 129 to 600 in this example. Plotting the metrics in order to achieve the best architecture is currently done by hand. In order to enhance this metrics exploration, we are working on an automatic metric exploration tool based on Integer Linear Programming.

Figure 7 represents the number of structures to be controlled (memory and network) in the generated architecture, when the user enable or disable the exploration of FIFO or LIFO structures for different input parallelism (1, 5 or 8 in this example).

Thus, when disabling FIFO (LIFO charts), the number of structures to be controlled is greater than when disabling the exploration of LIFO structures: this interleaving law used in this experiment is better suited for FIFO structure binding. The designer may give to FIFO related metrics a greater importance than the LIFO related ones.

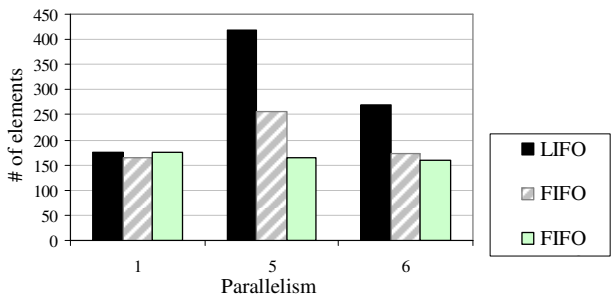

Figure 7: Enabling LIFO / FIFO binding

Moreover, Figure 7, also shows that enabling the FIFO and LIFO structures, is the best choice for most parallelism, for this interleaving law.

Finally, depending on designer targets, he may chose different metric settings, in order to reduce the number of memory points or the controller complexity.

4.2.2 STMicroelectronics interleaver design

Currently, we generate the different modes separately, while the reference design (from STMicroelectronics) integrates the three modes in a single 2400 memory points design. But when we concatenate our three designs (one for each mode) in a single architecture, the total area is about 14% smaller than the reference design. Future works will enable the generation of optimized multi-modes architectures to further reduce the area.

## 5. CONCLUSION

In this paper, we present a design space exploration methodology for Space-Time AdapteR STAR components. This approach relies on the formal modeling of communication constraints based on a Resource Compatibility Graph RCG describing timing relations between data. This methodology has been applied to interleaver design space exploration. Experimental results in the telecom domain have demonstrated the interest of this methodology. Formal modeling allows RTL architectures to be synthesized from a single C functional specification and under various I/O timing and parallelism constraints. We also show that with our methodology, the design space exploration is performed through metrics exploration. This allows enhancements based on refinements.

Future works will focus on the formal transformation of the RCG in order to generate multi-configuration and pipelined architectures. Moreover, we also investigate automatic plotting of the metrics by means of Integer Linear Programming methodologies